\input{aipcheck}

\documentclass[
    ,final            % use final for the camera ready runs
  ]
  {aipproc}

\usepackage{epsf,epsfig}
\layoutstyle{6x9}

\newcommand{\gtrsim}{\mathrel{\raisebox{0.4ex}{\hbox{$>$}}\kern-0.75em\raisebox{-0.5ex}{\hbox{$\sim$}}}}
\newcommand{\lesssim}{\mathrel{\raisebox{0.4ex}{\hbox{$<$}}\kern-0.75em\raisebox{-0.5ex}{\hbox{$\sim$}}}}
\newcommand{\psim}{\mathrel{\raisebox{0.4ex}{\hbox{$\propto$}}\kern-0.75em\raisebox{-0.5ex}{\hbox{$\sim$}}}}
\newcommand{\gsim}{\mathrel{\raisebox{0.4ex}{\hbox{$>$}}\kern-0.75em\raisebox{-0.5ex}{\hbox{$\sim$}}}}
\newcommand{\lsim}{\mathrel{\raisebox{0.4ex}{\hbox{$<$}}\kern-0.75em\raisebox{-0.5ex}{\hbox{$\sim$}}}}
\def\beq{\begin{equation}}
\def\eeq{\end{equation}}
\def\ba{\begin{array}}
\def\ea{\end{array}}

\begin{document}

\title{High-Energy Cosmic Rays from Galactic and Extragalactic 
Gamma-Ray Bursts}

\author{S.~D.~Wick}{
  address={NRL/NRC Research Associate, Code 7653, NRL, Washington, D.C. 20375-5352}
}

\author{C.~D.~Dermer}{
  address={Code 7653, Naval Research Laboratory, Washington, D.C. 20375-5352}
}

\author{A.~Atoyan}{
  address={CRM, Universit\'e de Montr\'eal, Montr\'eal, Canada H3C 3J7}
}

\begin{abstract}
A model for high-energy ($\gtrsim 10^{14}$~eV) cosmic rays (HECRs) 
from galactic and extragalactic gamma-ray bursts (GRBs) is summarized.
Relativistic outflows in GRBs are assumed to inject
power-law distributions of CR protons and ions to the
highest ($\gtrsim 10^{20}$ eV) energies.   
A diffusive propagation model for HECRs from a single recent GRB
within $\approx 1$~kpc from Earth explains the CR spectrum near and
above the knee.  
The CR spectrum at energies above $\sim 10^{18}$~eV  
is fit with a component from extragalactic GRBs. 
By normalizing the 
energy injection rate to that required to produce the CR flux 
from extragalactic sources
observed locally, we determine the amount of energy a typical 
GRB must release in the form of nonthermal hadrons.
Our interpretation of the HECR spectrum requires that 
GRBs are hadronically dominated, which would be confirmed
by the detection of HE neutrinos from GRBs.
\end{abstract}

\maketitle

%%%%%%%%%%%%%%%%%%%%%%%%%%%%%%%%%%%%%%%%%%%%
%% MAINMATTER
%%%%%%%%%%%%%%%%%%%%%%%%%%%%%%%%%%%%%%%%%%%%

\section{Introduction}

We summarize our model for HECRs, which
assumes that HECRs are accelerated in the relativistic 
shocks in GRBs \cite{wda03}.  GRBs inject HECRs with 
a single power-law from a low-energy cutoff $E_{min}\approx 10^{14}$~eV 
to a high-energy cutoff $E_{max}\gtrsim 10^{20}$~eV. 
We extend previous hypotheses \cite{vie95,wax95} that UHECRs originate from 
GRBs to include HECR origin from GRBs within our Galaxy \cite{der02}. 

GRBs are located in the disks of active star-forming galaxies 
such as the Milky Way.  
HECRs with energies $\lesssim 10^{18}$ eV diffuse
through their host galaxy.  UHECRs with energies $\gtrsim 10^{18}$~eV  
escape from their host galaxy and propagate almost rectilinearly in
extragalactic space.  UHECRs travelling over cosmological distances
have their spectrum modified by energy losses.  
An observer in the Milky Way will measure a superposition of UHECRs 
from extragalactic GRBs and HECRs  produced in our Galaxy.

By fitting to the measured KASCADE \cite{kam01} spectra of HECRs in the knee
region, we determine the properties of a Galactic GRB that produces
CRs.  Our fits to the UHECR spectrum \cite{hires} measured 
with the High-Res experiment imply that
GRBs must be strongly baryon-loaded, implying a detectable number of
high energy neutrinos with a km-scale neutrino detector such as
IceCube.

\section{The Model}

Our model of diffusive propagation of HECRs from a 
single nearby GRB in the Galaxy 
assumes pitch-angle scattering of CR trajectories in
Galactic magnetic fields on which is superposed a spectrum of
magneto-hydrodynamic (MHD) turbulence described by a distribution in
wave-number $k$ \cite{wda03}.
The Larmor radius of a CR propagating in a magnetic field of
strength $B = B_{\mu{\rm G}}$~$\mu$G is
$r_{\rm L} \cong A \gamma_6/(Z B_{\mu{\rm G}})$~pc,
where $\gamma= 10^{6}\gamma_6$ is the Lorentz factor of 
a CR with atomic (mass, charge)$=(A,Z)$.
The model assumes that the mean-free-path $\lambda$ between pitch-angle scatterings 
of a CR with Larmor radius $r_L$ is inversely proportional to 
the energy density in the MHD spectrum at wave-number $k\sim r_L^{-1}.$ 
A two component turbulence spectrum is assumed with wave-number index
$q = 5/3$ for a Kolmogorov-type (for large wave-number) and $q = 3/2$
for a Kraichnan-type (for small wave-number) turbulence.  The two
components give an energy-dependent break $\lambda$ at energy $E_Z
({\rm PeV})\cong Z B_{\mu{\rm G}}b_{pc},$ where $B_{\mu{\rm G}}=3$ and
$b_{pc}=1.6$ is the wavelength in parsecs of the MHD waves where the spectrum
changes from Kraichnan to Kolmogorov turbulence.

The diffusion radius $r_{dif} \cong 2\sqrt{\lambda c t/3}.$  When
$r \ll r_{dif},$ the number density of HECRs 
$ n(\gamma; r,t) \propto t^{-3/2}\times\gamma^{-p-1/2(3/4)}$
for $q = 5/3~(3/2).$  The measured spectrum is steepened
by ${3\over 2}(2-q)$ units because the diffusion coefficient
$D\propto\lambda \propto \gamma^{2-q}$ for an impulsive source \cite{aav95}.  
An injection spectrum with $p = 2.2$ gives
a measured spectrum $n_{Z,A}(\gamma; r,t)\propto
\gamma^{-s}$, with $s = 2.7$ at $E\ll E_Z$ and $s = 2.95$ at $E \gg
E_Z$. Because these indices are similar to the measured CR indices
above and below the knee energy, we adopt this model for CR
transport and investigate the implications of 
an injection spectrum $p = 2.2.$

Typical GRBs are thought to be beamed by a factor $500$ \cite{fra01}, meaning that
they are $1/500\times$ as energetic as observations imply and $500\times$
more frequent.  Accounting for the beaming factor, star formation
rate (SFR) evolution, and that
dirty and clean fireball transients \cite{bd00} may not be detected as GRBs,
the GRB rate per $L^*$ galaxy is $(1-3)\times 10^{-4}\;\;L^{* -1}{\rm ~yr}^{-1}.$
In the Milky Way we expect one GRB every 3,000 - 10,000 years with a high probability
that it will be beamed away from Earth.

%%%%ExtraGalactic:

UHECRs produced by extragalactic GRBs 
lose energy from momentum red shifting and photo-pair and photo-pion
production on the CMBR during propagation through a $\Lambda$-cold dark matter 
universe ($\Omega_{matter}=0.3$ and $\Omega_{\Lambda}=0.7$).  
Attenuation produces features in the UHECR
flux at characteristic energies 
$\sim 4\times 10^{18}$~eV and $\sim 5\times
10^{19}$~eV for photo-pair and photo-pion, respectively,
from distant sources ($z\gtrsim 1$).  We take the
local GRB CR luminosity density to be $\dot\varepsilon_{CR} =
f_{CR}\dot\varepsilon_{GRB,X/\gamma}$ where
$\dot\varepsilon_{GRB,X/\gamma}=10^{44}
~\rm{erg}~\rm{Mpc}^{-3}\rm{yr^{-1}}$ \cite{der02,vmg03}
 and $f_{CR}$ is the nonthermal baryon-loading factor required of
 the model.

The GRB cosmic rate-density evolution is assumed to follow the 
SFR history derived from the blue and UV luminosity
density of distant galaxies (see \cite{wda03}).  To accommodate
uncertainty in the SFR evolution we take two models, one based on
optical/UV measurements without extinction corrections (lower SFR), and
the other with extinction corrections \cite{Blain99}
(upper SFR).  The upper SFR is roughly a factor of 3(10) greater than
the lower SFR at red-shift $z=1(2).$ For $>10^{20}$~eV CRs, both
evolution models give the same flux, but the upper SFR contributes a
factor $\sim 3$ more CR flux over the lower SFR at energies $\lesssim
10^{18}$~eV.

\section{Results}

%%%%%%%%%%%%%%%%%
\begin{figure}
\centerline{\epsfig{file=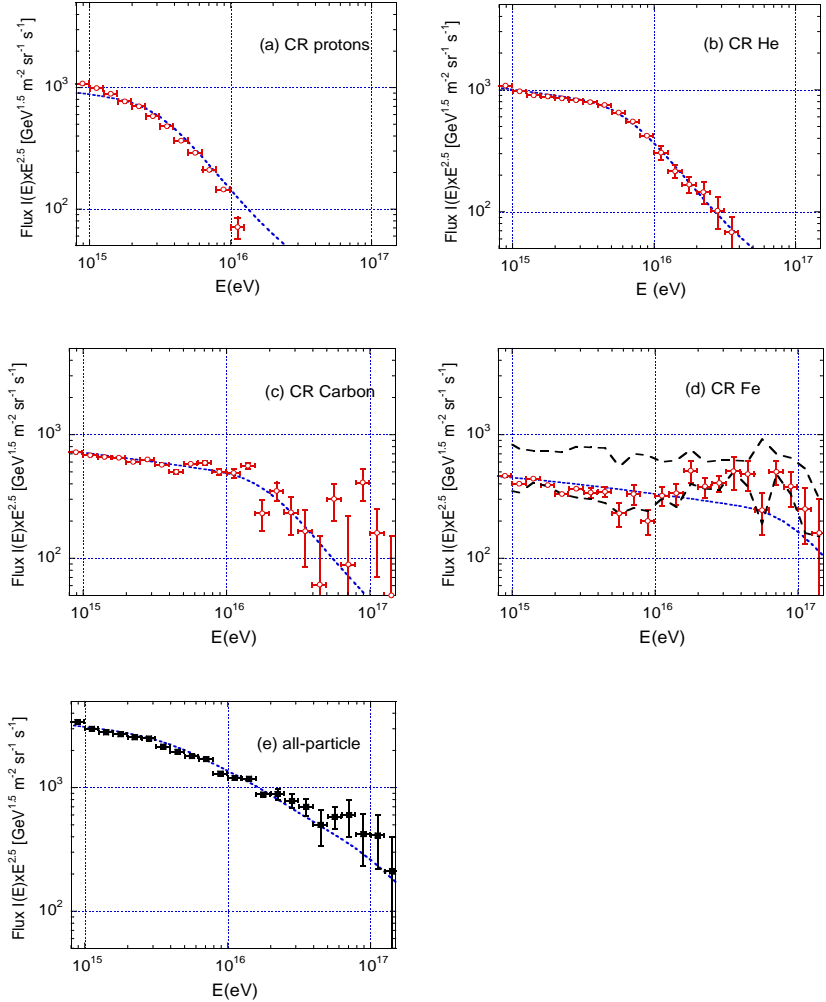,width=13.0cm,height=16.0cm }
%\vspace{6.5 in}
%\epsfxsize=200pt \epsbox{wda_fig4.eps}
}
\caption{Data points show preliminary KASCADE measurements of the CR proton (panel a),  He (panel b),
Carbon (panel c), Fe (panel d), and the all-particle spectrum (panel e), along
with model fits (dotted curves) to the CR ionic fluxes.  
In the model, a GRB that occurred $2.1\times 10^5$ years ago and at a distance of 500 pc
injects $10^{52}$ ergs in CRs.
The CRs isotropically diffuse via pitch-angle scattering
with an energy-dependent mean-free-path $\lambda$ in an MHD turbulence field. 
% given by Fig.~\ref{fig:kwk}. 
}
\label{fig:KASCADE}
\end{figure}
%%%%%%%%%%%%%%%%%

The preliminary (2001) KASCADE data \cite{kam01} are fit in
Fig.~\ref{fig:KASCADE} using the diffusion model described above 
with a GRB source
a distance $r\approx 500$ pc away that exploded $\approx 2\times 10^5$ yrs ago.
The break in the CR particle spectrum $\propto$ ionic charge $ Z$ is
apparent in the plots.  We did not perform a rigorous fit to the data,
but instead adjusted the wavenumber $k_1$ (where the turbulence
spectrum breaks) and the compositions of the ionic species
until a reasonable fit was obtained. The best fits  were obtained with
$k_1 \cong 1/1.6$~pc and composition  enhancements by a factor of 50 and 20 for C
and Fe, respectively, over Solar photospheric abundances.  The likelihood for
such an event is reasonable, and the corresponding anisotropy of the CRs
from a single source is shown to be consistent with observations \cite{wda03}.

%%%%%%%%%%%%%%%%%
\begin{figure}
\centerline{\epsfig{file=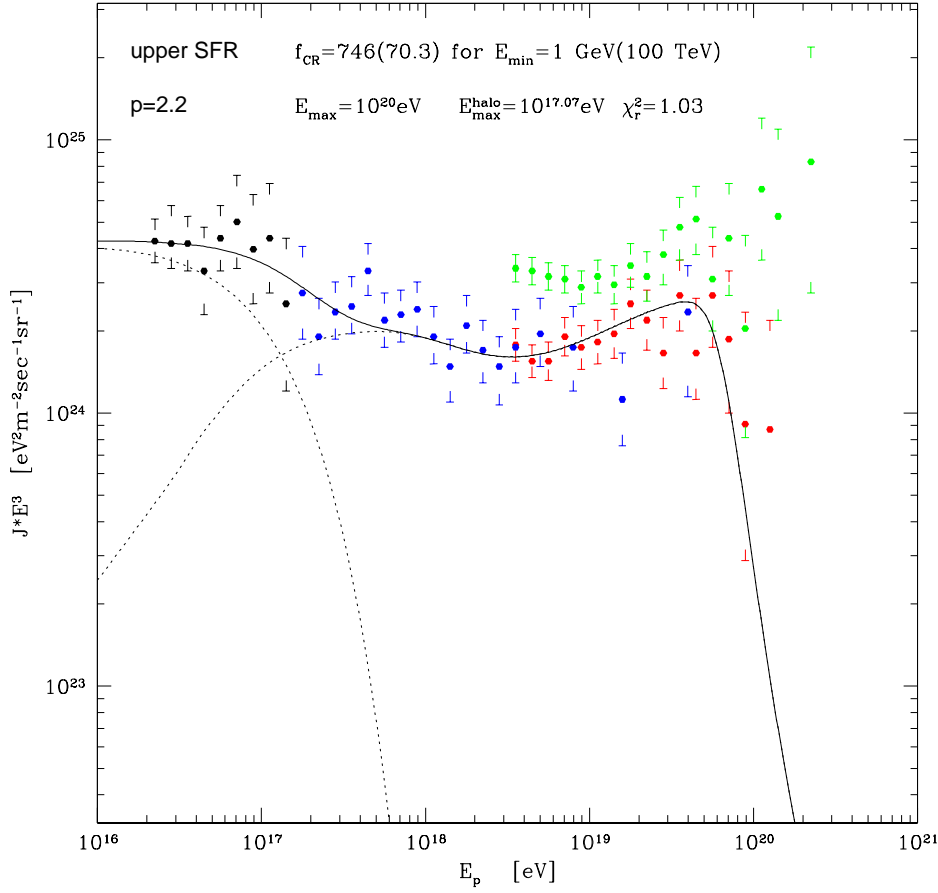,width=13.0cm,height=13.0cm}
%\centerline{
%\vspace{6.5 in}
%\epsfxsize=400pt \epsfbox{wda_fig16_clr.eps}
}
\caption{
Best fit to the KASCADE (black points), HiRes-I Monocular (red points),
and HiRes-II Monocular (blue points) data assuming a spectral cutoff
at the source of $E_{max}=10^{20}$~eV and using the upper limit to the SFR evolution.  
We also show the AGASA data \cite{tak98} (green points)
but do not include these in our fits.  The cutoff energy for the halo 
component is $E_{max}^{halo}=10^{17.07}$~eV and $\chi^{2}_r =1.03.$  
The requisite baryon loading factor is $f_{CR}=746(70.3)$ for  
a low energy cutoff at the source of $E_{min}=10^{9}(10^{14})$~eV.
This fit
implies that the transition from Galactic to extragalactic CRs occurs
near the second knee ($10^{17.6}$~eV) 
and that the ankle ($10^{18.5}$~eV) is associated with photo-pair
production.  
}
\label{fig:cr4}
\end{figure}
%%%%%%%%%%%%%%%%%

The combined KASCADE, HiRes-I and HiRes-II Monocular data
between  $\approx 2\times 10^{16}$~eV to $3\times 10^{20}$~eV are fit
in Fig.\ 2.
We investigated 8 separate cases  
\cite{wda03} with cutoff energy
$E_{\rm{max}} = 10^{20}$~eV and $10^{21}$~eV,
spectral indices of $p=2.0$ and $p=2.2$ (for optimizing KASCADE fits), 
and upper vs.~lower SFR evolution.  The free parameters we vary are
the galactic-halo-cutoff $E^{halo}_{max}$,
the baryon loading $f_{CR}$, 
and the intensity of the galactic halo CR component.
We give values of $f_{CR}$ corresponding 
to $E_{min}=10^{9}$~eV and $10^{14}$~eV. 
Fig.~\ref{fig:cr4} shows our best case, with $p=2.2$,
$E_{max}=10^{20}$~eV, and the upper SFR.  The $p = 2.0$ spectrum
provides a worse fit than the $p = 2.2$ case, although the CR energy
demand is less in this case because CRs are injected equally per unit
decade in particle energy.  The transition between galactic and
extragalactic CRs is found in the vicinity of the second knee
($10^{17.6}$~eV), consistent with a heavy-to-light composition change
\cite{fly93}.  The ankle ($10^{18.5}$~eV) is interpreted as a
suppression from photo-pair losses, analogous to the GZK suppression.

Our results imply that GRB blast waves are baryon-loaded by a factor
$f_{CR}\gtrsim 60$ compared to the energy injected and emitted by the
primary electrons that is inferred from hard X-ray and soft
$\gamma$-ray measurements of GRBs.  For the large baryon load
required for this model, calculations show that 100 TeV -- 100 PeV
neutrinos could be detected several times per year from all GRBs with
kilometer-scale neutrino detectors such as IceCube \cite{da03,wda03}.
Detection of even 1 or 2 neutrinos from GRBs with IceCube or a 
northern hemisphere neutrino detector would unambiguously demonstrate 
the high nonthermal baryon load in GRBs, and would provide compelling
support for this scenario for the origin of cosmic rays.

%%%%%%%%%%%%%%%%%%%%%%%%%%%%%%%%%%%%%%%%%%%%%%%%
%% BACKMATTER
%%%%%%%%%%%%%%%%%%%%%%%%%%%%%%%%%%%%%%%%%%%%%%%%

\vspace{5mm}

\noindent
The work of S.D.W.~was performed while he held a National Research
Council Research Associateship Award at the Naval Research Laboratory
(Washington, D.C).  The work of C.D.D.\ is supported by the Office of
Naval Research and NASA {\it GLAST} science investigation grant DPR \#
S-15634-Y. ~A.A.\ acknowledges support and hospitality during visits
to the High Energy Space Environment Branch.

%% If the bibliography is produced without BibTeX comment out the

%%\bibliographystyle{aipprocl} % if natbib is missing


%\bibliography{sample}

%\section{References}

%  \typeout{******************************************}
%  \typeout{** Please run "bibtex \jobname" to optain}
%  \typeout{** the bibliography and then re-run LaTeX}
%  \typeout{** twice to fix the references!}
%  \typeout{******************************************}

\end{document}